# Unidirectional orbital magnetoresistance in light metal/ferromagnet bilayers

Shilei Ding[*], Paul Noël, Gunasheel Kauwtilyaa Krishnaswamy, Pietro Gambardella[*]

*Department of Materials, ETH Zürich, 8093 Zürich, Switzerland*

**We report the observation of a unidirectional magnetoresistance (UMR) that originates from the nonequilibrium orbital momentum induced by an electric current in a naturally oxidized Cu/Co bilayer. The orbital-UMR scales with the torque efficiency due to the orbital Rashba-Edelstein effect upon changing the Co thickness and temperature, reflecting their common origin. We attribute the UMR to orbital-dependent electron scattering and orbital-to-spin conversion in the ferromagnetic layer. In contrast to the spin-current induced UMR, the magnon contribution to the orbital-UMR is absent in thin Co layers, which we ascribe to the lack of coupling between low energy magnons and orbital current. The magnon contribution to the UMR emerges in Co layers thicker than about 5 nm, which is comparable to the orbital-to-spin conversion length. Our results provide insight into orbital-to-spin momentum transfer processes relevant for the optimization of spintronic devices based on light metals and orbital transport.**

The generation of nonequilibrium angular momentum is essential to the functioning of spintronic devices [1]. Various mechanisms based on spin-orbit coupling (SOC) have been proposed for the generation of spin currents, such as the spin Hall effect [2,3], Rashba-Edelstein effect [4,5], and spin-momentum locking in topological insulators [6,7]. Besides allowing for the electrical manipulation of magnetism, including magnetization switching [8,9], domain wall motion [10,11,12], and magnon excitation [13,14], spin currents strongly affect the electrical conductivity of heterostructures, resulting in the giant magnetoresistance (GMR) [15], spin Hall magnetoresistance (SMR) [16,17], and unidirectional magnetoresistance (UMR) [18-33]. The UMR is a nonreciprocal resistive effect that arises in nonmagnetic/ferromagnetic metal bilayers due to the interaction of an electrically-induced spin current with the magnetization [18]. Unlike the anisotropic and spin Hall magnetoresistance, the UMR is proportional to the magnitude of the electric current and changes sign upon reversal of either current or magnetization [18]. For this reason, it can be used to detect magnetization switching using simple two-terminal resistance measurements in planar devices [33,34]. Two distinct mechanisms have been shown to contribute to the UMR [23]. One is the interfacial and bulk spin-dependent scattering [18-23], whereby the resistance is modulated by the current-induced spin accumulation at the interface or in the bulk of the ferromagnet, similar to the giant magnetoresistance [35]. The other is electron-magnon scattering, which increases (decreases) the resistance upon the excitation (annihilation) of magnons induced by the spin current in the ferromagnet [24-31,36,37]. Recently, the UMR was reported also in systems lacking strong spin-orbit coupling, such as in surface oxidized Cu*/NiFe bilayers, and attributed to the vorticity of the electric current in a system in which the electronic mobility varies with thickness [38]. The UMR thus provides fundamental insight into the interaction of an angular momentum current with a magnetic system and a telltale signature of charge-to-spin conversion in heavy metals, semiconductors, and topological materials.

Theoretical work predicts a new approach for the realization of spin-orbitronic devices based on electrically-induced orbital currents in light metal/ferromagnetic metal (FM) bilayers [39-43]. Independently of SOC, inversion symmetry breaking in such heterostructures is sufficient for the emergence of a nonequilibrium orbital angular momentum upon application of an electric field, as exemplified by the orbital Hall and orbital Rashhba Edelstein effect [39-41]. Recent experiments provide evidence for large orbital torque and orbital Rashba-Edelstein magnetoresistance in light metal/FM systems [44-50], supporting the idea that the orbital torque can be as efficient as the spin torque. Interestingly, however, nonequilibrium orbital and spin angular momenta interact with the local magnetization in fundamentally different ways. Orbital-to-spin conversion is required to generate an orbital torque [48,50], whereas the effective diffusion length of the orbital current is quite long in contrast to the spin current [49,50]. Up to now, it is unknown if an orbital analog to the UMR exists, which we call orbital-UMR, and if it has properties similar to the spin-current induced UMR.

In this paper, we report evidence for the orbital-UMR in naturally oxidized Cu (denoted by Cu* hereafter)/Co bilayers without heavy elements with large SOC. We find that the orbital-UMR shares the same symmetry with the spin-UMR and that the orbital torque efficiency and orbital-UMR vary simultaneously by changing the thickness of the FM layer and temperature, reflecting a similar origin. From the magnetic field dependence, we conclude that electron-magnon scattering does not play a role in the orbital-UMR in thin Co layers due to the lack of net spin current generation, unlike in the spin-UMR. The orbital-UMR in Cu*/Co is thus mainly attributed to the alteration of the resistance through the orbital angular momentum transport and orbital-to-spin conversion. The magnon contribution to the UMR emerges in Co layers thicker than 5 nm, which provides an estimate for the length scale of orbital-to-spin conversion in a ferromagnetic metal.

The Cu*/Co samples were grown on the Si/SiO$_2$ substrates by dc magnetron sputtering with an Ar pressure of $4.3 \times 10^{-3}$ mbar. Double Hall bar devices with a width of 10 μm and aspect ratio of ~1 were patterned by photolithography and lift-off, and the samples were stored in air for 2 days to naturally oxidize before the transport measurements. The single layer of 3 nm Cu* is electrically insulating (see Supplementary Note 1 [51]). The magnetotransport measurements were carried out in a cryogenic system and at room temperature using an alternate (ac) excitation current with a frequency of 10 Hz. The first and second harmonic longitudinal and Hall resistance were subsequently analyzed [18].

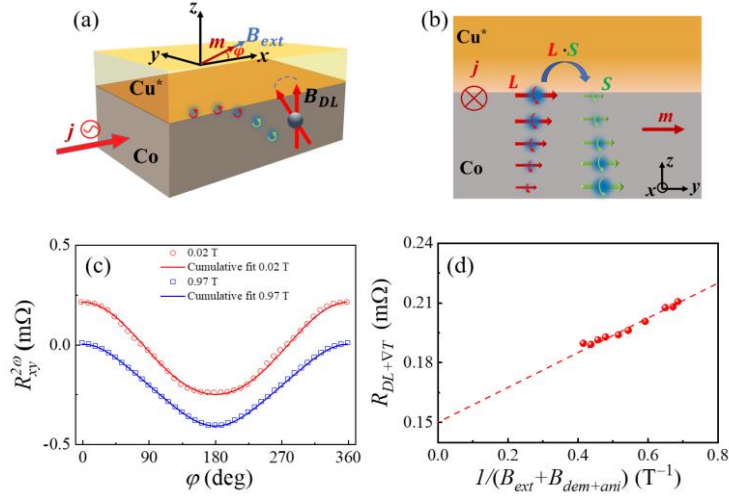

FIG. 1 (a) Schematic illustration of the orbital torque. The coordinate axes, angles, and measurement parameters are also shown. (b) Schematic of the conversion of an orbital current into a spin current, the red and green arrows represent the orbital ($L$) and spin ($S$) angular momenta. The source of $L$ and $S$ are the orbital Rashba-Edelstein effect and orbital-to-spin conversion, respectively. (c) Transverse second-harmonic resistance, $R_{xy}^{2\omega}$, of Cu*(3)/Co(2.5) (thickness in nanometers) as a function of the angle $\varphi$ between the magnetic field $B_{ext}$ and current direction measured for $B_{ext} = 0.02$ and $0.97$ T. (d) $R_{DL+\nabla T}$ as a function of the inverse effective field $1/(B_{ext} + B_{dem+ani})$. The dashed line is a linear fit according to Eq. (2).

We first demonstrate the presence of current-induced torque in Cu*/Co. A schematic of the sample and the generation of orbital and spin momenta are shown in Fig. 1(a) and (b), respectively. The current-induced torque can be extracted by measuring the in-plane angular dependence of the transverse second-harmonic resistance $R_{xy}^{2\omega}$ expressed as [58,59]

$$R_{xy}^{2\omega}(\varphi) = R_{DL+\nabla T}\cos(\varphi) + R_{PHE}(2\cos^3(\varphi) - \cos(\varphi))\frac{B_{FL} + B_{oe}}{B_{ext}} \quad (1)$$

$$R_{DL+\nabla T} = \frac{1}{2}R_{AHE}\frac{B_{DL}}{B_{ext} + B_{dem+ani}} + R_{\nabla T} \quad (2)$$

Here, $B_{DL}$, $B_{FL}$, and $B_{oe}$ represent the current-induced effective field from the damping-like torque, field-like torque, and Oersted field, $R_{AHE}$ and $R_{PHE}$ correspond to the anomalous Hall resistance and planar Hall resistance, respectively, and $R_{\nabla T}$ is the transverse resistance due to the Hall voltage $\sim (\nabla T \times m) \cdot y$ induced by the anomalous Nernst effect and, possibly, by the orbital analog of the spin Seebeck effect for a thermal gradient $\nabla T$ perpendicular to the magnetization $m$. $B_{ext}$ is the applied magnetic field and $B_{dem+ani}$ stands for the effective demagnetization and anisotropy field, which we estimate from the anomalous Hall effect (see Supplementary Note 3 [51]). Figure 1(c) shows the angular dependent $R_{xy}^{2\omega}$ measured at 280 K with $B_{ext} = 0.02$ and 0.97 T for an ac current of 8 mA (peak value). Fitting these curves by Eq. (1) allows us to find the coefficients $R_{DL+\nabla T}$. We further measured $R_{xy}^{2\omega}$ at various $B_{ext}$ to separate the torque and thermal contributions to $R_{DL+\nabla T}$ [40]. According to Eq. (2), $R_{DL+\nabla T}$ scales proportionally to $1/(B_{ext} + B_{dem+ani})$ [Fig. 1(d)] and a linear fit gives $R_{\nabla T} = 0.15 \pm 0.01$ mΩ, and $\frac{1}{2}R_{AHE}B_{DL} = 0.079 \pm$ 0.006 mΩ·T. Using $R_{AHE} = 0.161 \pm 0.005$ Ω (see Supplementary Note 3 [51]), we obtain the effective field $B_{DL} = 0.98 \pm 0.04$ mT corresponding to the damping-like orbital torque. Finally, by measuring the variation of $B_{DL}$ as a function of applied current (see Supplementary Note 4 [51]), we calculate the torque efficiency per unit applied current density $\xi_{DL} = \frac{2e}{\hbar} M_s t_{Co} B_{DL}/j$ [10], where $M_s$ represents the saturation magnetization. The torque efficiency for Cu*(3)/Co(2.5) at 280 K is thus determined to be $0.011 \pm 0.001$ (see Supplementary Note 4 [51]). We remark that the current-induced torque in Cu*/Co is attributed to the orbital Rashba-Edelstein effect, whereby the strong built-in electric field from the oxygen gradient is crucial in the absence of strong SOC [41,49]. In such a case, an electric current flowing at the Cu*/Co interface generates orbital angular momentum, which diffuses into the adjacent ferromagnetic layer, where the SOC converts it to a spin current and thus to an orbital torque [Fig. 1(a,b) and Refs. 45-48]. The presence of a self-induced torque due to the spin-polarized current flowing in Co is excluded in our samples by control measurements on a single Co(5) layer without Cu* capping and on a Cu(7)/Co(5) bilayer (see Supplementary Note 5 [51]).

We now turn to explore the UMR in the Cu*/Co system. Because the UMR is a nonlinear resistance proportional to the current, it emerges as a second-harmonic contribution to the longitudinal resistance $R_{xx}^{2\omega}$. The angular dependence of $R_{xx}^{2\omega}$ is given by [18, 30]

$$R_{xx}^{2\omega}(\varphi) = R^* \cdot \sin(\varphi) - 2\Delta R_{xx}^{1\omega} \frac{B_{FL}+B_{oe}}{B_{ext}} \cos^2(\varphi) \sin(\varphi), \qquad (3)$$

where $\Delta R_{xx}^{1\omega}$ is the change of the first harmonic resistance. $R^* = gR_{\nabla T} + R_{UMR}$ is the longitudinal magnetoresistance that includes the contribution from the thermal voltage $\sim (\nabla T \times \boldsymbol{m}) \cdot \boldsymbol{x}$ and the UMR $\sim \boldsymbol{j} \times \boldsymbol{m}$. The former is the same effect that contributes to Eq. (1) rescaled by the geometric factor $g = l/w$, where $l$ and $w$ indicate the length and the width of the Hall bar [18]. As we are interested in the first term of Eq. (3), we measured the angular dependence of $R_{xx}^{2\omega}$ at large field, as shown in Fig. 2(a). The experimental $R_{xx}^{2\omega}(\varphi)$ can be well fitted with Eq.(3), which gives $R^* = 0.303 \pm 0.008$ mΩ. Measuring $R^*$ as a function of field shows the characteristic dependence expected of both UMR and thermal voltage [18], with $R_{xx}^{2\omega}$ changing sign when the direction of $\boldsymbol{m}$ reversal along $y$, as shown in Fig. 2(b). To separate the two contributions to $R^*$ we estimate $g$ from the ratio of the longitudinal resistance to the planar Hall effect, $\frac{\Delta R_{xx}^{1\omega}}{\Delta R_{xy}^{1\omega}} \approx 1.2$ (see Supplementary Note 6 [51]), which is more precise than using the nominal ratio $\frac{l}{w} = 1$ that is affected by the resolution of the lithography process. Using $R_{\nabla T}$ obtained from the torque measurements discussed above, which is consistent with the value obtained via the field-dependent measurement of $R_{xy}^{2\omega}$ along the x-axis (see Supplementary Note 7 [51]), we estimate $\frac{gR_{\nabla T}}{R^*} = 60\%$.

The additional magnetoresistance $R^* - gR_{\nabla T}$ is therefore attributed to the UMR in the Cu*/Co system, where $R_{UMR} = 0.12 \pm 0.01$ mΩ. We note that a similar UMR was reported in Cu*/NiFe bilayers and attributed to the inhomogeneous current flow due to the strong mobility gradient inside naturally oxidized Cu and spin-vorticity coupling [38]. However, this interpretation is at variance with the increasing body of evidence supporting the emergence of orbital Hall and Rashba-Edelstein effect in light metal systems, including metals with no mobility gradient [44-50]. Moreover, the spin-vorticity effect cannot account for the dependence of the orbital torque on the type and thickness of the ferromagnetic layer [45]. Given the absence of elements with strong SOC, the insulating character of Cu*, and the presence of the interfacial oxygen gradient in Cu*/Co, we thus ascribe the UMR to the orbital accumulation induced by the orbital Rashba-Edelstein effect [41,44,45,49]. The following orbital-UMR data are collected from the angle-dependent measurements of $R_{xx}^{2\omega}$.

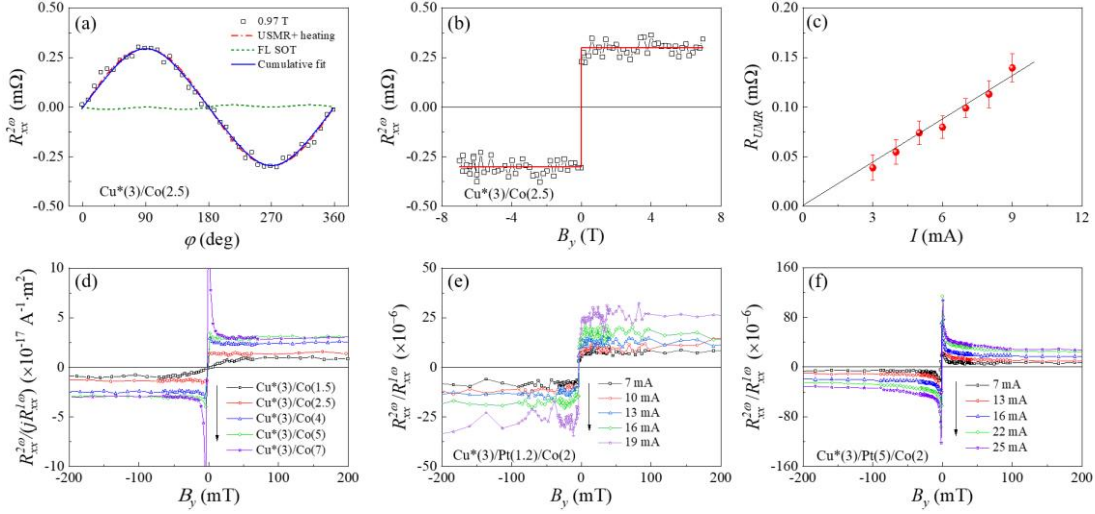

FIG. 2 (a) Longitudinal second-harmonic resistance, $R_{xx}^{2\omega}$, of Cu*(3)/Co(2.5) as a function of $\varphi$ at an applied field $B_{ext} = 0.97$ T. (b) Field dependence of $R_{xx}^{2\omega}$ for $B_{ext}$ applied parallel to the y-axis ($B_y$). The applied current is 8 mA (peak value) and the temperature is 280 K for the measurements in (a) and (b). (c) Variation of the orbital-UMR as a function of the applied current obtained from the angle-dependent measurements of $R_{xx}^{2\omega}$. The solid line is a linear fit to the data constrained through the origin. (d) Field dependence of $R_{xx}^{2\omega}$ in Cu*(3)/Co($t_{Co}$) as a function of Co thickness. The peak magnetoresistance asymmetry at a low field is the signature of the magnon-UMR, which appears for $t_{Co} > 4$ nm. The applied currents are 4, 8, 14, 15, and 18 mA. (e) Field dependence of $R_{xx}^{2\omega}$ in Cu*(3)/Pt(1.2)/Co(2) and (f) Cu*(3)/Pt(5)/Co(2) as a function of applied current.

We now discuss the properties of the UMR in Cu*/Co. Figure 2(c) shows the $R_{UMR}$ of Cu*(3)/Co(2.5) as a function of applied current $I$ at 280 K. As expected, the orbital-UMR at a high field is proportional to the current density, similar to the spin-UMR due to interfacial and bulk spin-dependent scattering. A striking difference between the orbital- and spin-UMR is the absence of the low-field enhancement of the UMR in Cu*/Co relative to Pt/Co and other systems based on heavy metals and topological insulators, which is due to the spin current exciting or annihilating magnons and electron-magnon scattering [23,24,36,37]. The field-dependent $R_{xx}^{2\omega}$ of Cu*(3)/Co(2.5) reported in Fig. 2(b) shows no sign of such an enhancement despite the large applied current density of $4 \times 10^{11}$ Am$^{-2}$. However, the magnon-induced contribution to the UMR emerges gradually in thicker Co samples and becomes prominent at $t_{Co} = 7$ nm, as evidenced by the low-field divergence of $R_{xx}^{2\omega}$ in Fig. 2(d). The dependence of $R_{UMR}$ on Co thickness is thus a distinctive feature of the orbital-UMR, which we attribute to the lack of interaction of the orbital current with magnons in the thinner Co layers and the conversion of the orbital current into a spin current in the thicker Co layers, as discussed further below. A similar conclusion on the orbital character of the UMR is reached when comparing the field dependence of $R_{xx}^{2\omega}$ in Cu*/Co with that of Cu*/Pt/Co, where the Pt spacer provides both efficient conversion of the orbital current injected from the Cu* interface [45] and the generation of a spin current due to the SHE [2,3]. Indeed, the low-field enhancement of the UMR is absent in Cu*(3)/Pt(1.2)/Co(2) [Fig. 2(e)], because Pt can efficiently generate a spin current and convert the orbital current from Cu* only on length scales larger than the spin diffusion length [45], and recovered in Cu*(3)/Pt(5)/Co(2), as shown in Fig. 2(f). This low-field enhancement gives rise to an inverse power law dependence of $R_{UMR}$ on $B_{ext}$ and a characteristic scaling of $R_{UMR}$ with current $I$ as $aI + bI^3$, which has been shown to be proportional to the magnon population in the ferromagnetic layer [25,23] (see also Supplementary Note 8 [51]). The strong nonlinearity of the magnon contribution with current explains its abrupt increase in Co films thicker than 5 nm seen in Fig. 2(d). For a constant current density, the current injected in thicker samples is larger than in thinner samples, which generates more heat and magnons, making the effect stronger in thicker Co. Moreover, the magnon UMR increases as the magnon stiffness of the ferromagnetic layer decreases, as observed in thicker Co films [23].

Our results indicate that an orbital current cannot directly excite or annihilate magnons, consistently with the fact that magnons are bosonic spin excitations. This observation underscores a key difference between the orbital- and spin-UMR. However, the question remains as to why the orbital current in the thinner ferromagnetic layers gives rise to the orbital torque discussed above but does not influence the magnon population in a significant way. Although recent work predicts the possibility of orbital-magnon coupling in special symmetry conditions [60], theoretical insight into the interaction between an orbital current and magnons is generally lacking. Here we offer a few considerations with the hope of stimulating further work in this direction. (i) An orbital current does not couple to magnons directly, because the magnon creation and annihilation operators are spin operators. However, an orbital current can directly affect the electrical conductivity of a ferromagnet owing to orbital-dependent electron scattering, e.g., due to orbital-selective *s-d* transitions [61]. Thus, the observed field and current dependence of the UMR of Cu*(3)/Co(2.5) could be explained by involving uniquely the current-induced orbital accumulation and orbital-dependent electron mobility of Co in analogy with the UMR due to spin-dependent scattering reported in spin systems [18-21]. The conversion of an orbital current into a spin current is not a necessary condition to induce the orbital-UMR. (ii) The small but finite orbital torque measured in the thinner Co layers may be due to the incipient orbital-to-spin conversion taking place in Co or to the direct coupling of the orbital current to the local spin magnetization due to spin-orbit coupling. In addition to these points, one might consider that (iii) the orbital-to-spin conversion can be more (less) efficient when the polarization of the orbital current is perpendicular (parallel) to the local magnetization, which corresponds to the geometry that determines the torque (UMR). This might be the case if the spin-orbit induced mixing of spin-up and spin-down states in 3*d* ferromagnets is smaller (larger) when the spins are aligned parallel (perpendicular) to the magnetization direction [62]. Moreover, (iv) interface effects also play a role in magnon excitation processes. The low-frequency magnons responsible for the spin-UMR can be excited directly by the spin current flowing across a heavy metal/ferromagnet interface via electron spin-flip scattering, at a rate proportional to the magnon density and the current-induced shift of the spin-dependent electrochemical potential at the interface [25]. In an orbital system, the light metals or oxides do not generate a sizeable spin current at the interface with the ferromagnetic layer, and the spin angular momentum is then generated locally in the ferromagnets via orbital-to-spin conversion. In such a case, the efficiency of the magnon excitation process would be considerably reduced. Overall, these considerations indicate that the orbital current generated in a light metal system can interact in different ways with the local magnetization compared to a spin current, which is not yet fully understood.

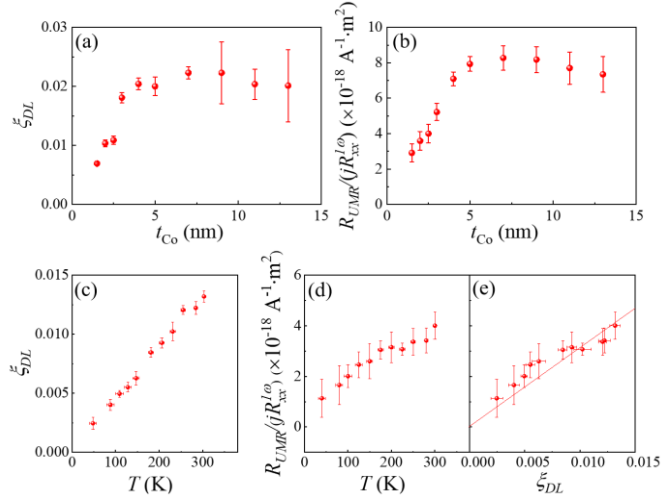

FIG. 3 (a) Orbital torque efficiency and (b) UMR of Cu*(3)/Co($t_{Co}$) as a function of Co thickness The data were collected at room temperature. (c) Temperature dependence of Orbital torque and (d) UMR of Cu*(3)/Co(2.5) as a function of temperature. The temperature scale has been calibrated by measuring the resistance change as a function of temperature. (e) Plot of the orbital-UMR vs orbital torque obtained from the data shown in (c,d). The line is a linear fit to the data constrained through the origin.

To gain further insight into the properties of the orbital-UMR, we measured the orbital torque efficiency $\xi_{DL}$ and $R_{UMR}$ (high field values) as a function of Co thickness ($t_{Co}$). Figure 3(a) shows $\xi_{DL}$ of Cu*(3)/Co($t_{Co}$): we observe that the torque efficiency increases up to $t_{Co} \approx 5$ nm, and then saturates or decreases slightly in thicker Co films. This dependence is consistent with the absorption and conversion of orbital angular momentum in Co, which occurs over an extended thickness range [48-50]. To compare the UMR in different samples, we consider the ratio $R_{UMR}/(jR_{xx}^{1\omega})$ [18], where $j$ is the current density obtained by dividing the total current by the cross section of the Co layer and $R_{UMR}$ is obtained at high field, thus excluding the magnon contribution, which does not scale linearly with current. Figure 3(b) shows that $R_{UMR}/(jR_{xx}^{1\omega})$ behaves similarly to $\xi_{DL}$ as a function of thickness, reaching a maximum value of about $9 \times 10^{-18}$ A$^{-1}$m$^2$ at $t_{Co} \approx 7$ nm. We note the torque efficiency of Cu*(3)/Co($t_{Co}$) is one order of magnitude magnitude smaller compared to Pt/Co, and the UMR per current density divided by the total resistance is also one order of magnitude smaller than the UMR in Pt/Co [21]. However, the magnitude of UMR per torque efficiency is similar for the orbital-UMR in Cu*/Co and the spin-UMR in Pt/Co.

We further discuss the temperature-dependent orbital-UMR in Cu*(3)/Co(2.5) bilayers. Figures 3(c) and (d) show that both $\xi_{DL}$ and $R_{UMR}/(jR_{xx}^{1\omega})$ decrease monotonically as a function of temperature. Previous measurements of the spin-UMR have shown that also the spin-dependent scattering and magnon contributions to the UMR in Pt/Co decrease monotonically upon lowering the temperature (Supplementary Information of Ref. [23]). In the thinner Cu*/Co($t_{Co}$) layers, the magnon contribution is absent, and we conclude that either current shunting through the Co layer decreases the current flow through Cu* or the orbital current generation and/or the orbital-to-spin conversion is less efficient at low temperatures. This conclusion is consistent with the nearly linear correlation observed between $\xi_{DL}$ and $R_{UMR}/(jR_{xx}^{1\omega})$ [Fig. 3(e)].

In summary, we reported evidence of a unidirectional magnetoresistance in a light metal system originating from an orbital current, which we refer to as the orbital-UMR. By varying the thickness of the ferromagnetic layer and temperature, we find that the magnitude of the orbital-UMR scales linearly with the orbital torque efficiency, supporting a common origin for the two effects. In contrast to the spin-UMR, the magnon contribution to the resistance asymmetry is absent in the orbital-UMR of thin ferromagnetic layers, which shows that orbital currents do not couple directly to magnons. Following the generation of nonequilibrium orbital moments at the Cu*/Co interface, the orbital-UMR is thus ascribed to orbital-to-spin conversion and possibly to orbital-dependent electron scattering in the ferromagnetic Co layer. The emergence of the magnon enhancement of the UMR in thicker Co layers provides evidence of the conversion of an orbital current into a spin current, which occurs on a length scale of 5 nm. Our findings not only demonstrate the current-induced UMR in a light metal system based on the orbital Rashba-Edelstein effect but also pave the way toward the microscopic understanding of how orbital angular momentum interacts with the local magnetization.

**Acknowledgments**
This work was funded by the Swiss National Science Foundation (Grant No. 200020_200465). P.N. acknowledges the support of the ETH Zurich Postdoctoral Fellowship Programme 19-2 FEL-61.